\documentclass[aps,prd,twocolumn,showpacs,showkeys,nofootinbib,floatfix]{revtex4-1}

\usepackage{amssymb}
\usepackage{amsmath}
\usepackage{graphicx}
\usepackage{hyperref}

\begin{document}

\title{Constraints on the Holographic Dark Energy Model from Type Ia Supernovae, WMAP7, Baryon Acoustic Oscillation and Redshift-Space Distortion}

\author{Lixin Xu}
\email{lxxu@dlut.edu.cn}

\affiliation{Institute of Theoretical Physics, School of Physics \&
Optoelectronic Technology, Dalian University of Technology, Dalian,
116024, P. R. China}

\affiliation{College of Advanced Science \& Technology, 
Dalian University of Technology, Dalian, 116024, P. R. China}

\begin{abstract}
In this paper, we use the joint measurement of geometry and growth rate from matter density perturbations to constrain the holographic dark energy model. The geometry measurement includes type Ia supernovae (SN Ia) Union2.1, full information of cosmic microwave background (CMB) from WMAP-7yr and baryon acoustic oscillation (BAO). For the growth rate of matter density perturbations, the results $f(z)\sigma_8(z)$ measured from the redshift-space distortion (RSD) in the galaxy power spectrum are employed. Via the Markov Chain Monte Carlo method, we try to constrain the model parameters space. The jointed constraint shows that $c=0.750_{-    0.0999-    0.173-    0.226}^{+    0.0976+    0.215+    0.319}$ and $\sigma_8=0.763_{-    0.0465-    0.0826-    0.108}^{+    0.0477+    0.0910+    0.120}$ with $1,2,3\sigma$ regions. After marginalizing the other irrelevant model parameters, we show the evolution of the equation of state of HDE with respect to the redshift $z$. Though the current cosmic data points favor a phantom like HDE Universe for the mean values of the model parameters in the future, it can behave like quintessence in $3\sigma$ regions. 
\end{abstract}



\maketitle

\section{Introduction}

The holographic principle says that the number of degrees of freedom in a bounded system should be finite and has relations with the area of its boundary \cite{ref:holo0}. By applying the so-called holographic principle to cosmology, one derives a relation between vacuum density and a cosmological scale $\rho_{\Lambda}
=3c^2 M^{2}_{pl} L^{-2}$ \cite{ref:holo0,ref:holo1,ref:holo2}, where $c$ is a numerical constant and
$M_{pl}$ is the reduced Planck Mass $M^{-2}_{pl}=8 \pi G$. The obtained vacuum energy, dubbed as holographic dark energy (HDE), can push our Universe into an accelerated expansion phase at late time \cite{ref:Riess98,ref:Perlmuter99}.
By taking different cosmological scale, for example the Hubble
horizon \cite{ref:holo0,ref:holo1,ref:XuJCAP}, the event horizon or the particle horizon \cite{ref:holo2} as discussed by
\cite{ref:holo0,ref:holo1,ref:holo2} and the Ricci scalar \cite{ref:Ricci}, one has different HDE model. Based on the idea that gravity as an entropic force \cite{ref:Verlinde},  a similar DE density was given in \cite{ref:EFS} where a linear combination of $H^2$ and $\dot{H}$ was also presented, see also \cite{ref:Hcom1,ref:Hcom2}. Furthermore  generalized HDE models $\rho_{R}=3c^2M^{2}_{pl}Rf(H^2/R)$ and $\rho_{h}=3c^2M^{2}_{pl}H^2g(R/H^2)$ were also presented in Ref. \cite{ref:XuGHDE}. In this paper, we consider the {\it typical} HDE model where the future event horizon 
\begin{equation}
R_{eh}(a)=a\int^{\infty}_{t}\frac{dt^{'}}{a(t^{'})}=a\int^{\infty}_{a}\frac{da^{'}}{Ha^{'2}}\label{eq:EH}
\end{equation}
is taken as a large cosmological scale, i.e. the IR cut-off $L=R_{eh}(a)$. This horizon is the boundary of the volume a fixed observer may eventually observe. This model has been confronted by cosmic observations extensively \cite{ref:Kao2005,ref:Gong2005,ref:holoZhang}, for recent results, please see \cite{ref:holoXu} and \cite{ref:holowang2012}. In the literature, to the best of our knowledge, only the geometry information which includes the luminosity distance $d_L$ from SN Ia, the angular diameter distance $D_A$ from BAO and the full information of CMB from WMAP-7yr were used to constrain this model, for examples please see \cite{ref:holoXu} and \cite{ref:holowang2012}. As is well known, to discriminate the cosmological models the geometry information is not enough due to the degeneracies between model parameters. It means that different cosmological models can have the same background evolution history. However the dynamical evolution would be very different even if  they have the same background evolution. Which is to say the dynamical evolution is important to break the possible degeneracy.

Thanks to the measurement of the cosmic growth rate via the redshift-space distortion (RSD) which relates to the evolutionary speed of matter density contrast, now one can constrain the evolutions of the density contrast $\delta$ through $f(z)\sigma_8(z)$, where $f(z)=d\ln \delta/d \ln a$ is the growth rate of matter and $\sigma_8(z)$ is the rms amplitude of the density contrast at the comoving $8h^{-1}$ Mpc scale. Here $h$ is the normalized Hubble parameter $H_0=100h\text{km sec}^{-1}\text{Mpc}^{-1}$. Here we should notice that the growth rate of structure $f(z)$  has been used to constrain the dark energy model and to investigate the growth index in the literature, see \cite{ref:weakness} for examples. However, the observed values of the growth rate $f_{obs}=\beta b$ are derived from the redshift space distortion parameter $\beta(z)$ and the linear bias $b(z)$, where a particular fiducial $\Lambda$CDM model is used. It means that the current $f_{obs}$ data can only be used to test the consistency of $\Lambda$CDM model. This is the weak point of using $f_{obs}$ data points. Moreover, the measurements of the linear growth rate are degenerate with the bias $b$ or clustering amplitude in the power spectra. To remove this weakness, Song \& Percival proposed to use $f\sigma_8(z)$ which is almost model independent and provides good test to dark energy models even without the knowledge of the bias or $\sigma_8$ \cite{ref:Song}. Recently, the observed values of $f(z)\sigma_8(z)$ were provided by the 2dFGRS \cite{Percival04}, WiggleZ \cite{Blake}, SDSS LRG \cite{Samushia11}, BOSS \cite{Reid12}, and 6dFGRS \cite{Beutler11}. The latest RSD data points were also summarized in \cite{Samushia12}. For convenience, we show the data points used in this paper in Table \ref{table:fs8data}, see also Table 1 of Ref. \cite{Samushia12}.
\begin{center}
\begin{table}[tbh]
\begin{tabular}{ccl}
\hline \hline
$z$ &  $f\sigma_8(z)$ & Survey and Refs \\
\hline
0.067 & 0.42$\pm$0.06 & 6dFGRS (2012) \cite{Beutler11} \\
0.17 & 0.51$\pm$0.06 & 2dFGRS (2004) \cite{Percival04} \\
0.22 & 0.42$\pm$0.07 & WiggleZ (2011) \cite{Blake} \\
0.25 & 0.39$\pm$0.05 & SDSS LRG (2011) \cite{Samushia11} \\
0.37 & 0.43$\pm$0.04 & SDSS LRG (2011) \cite{Samushia11} \\
0.41 & 0.45$\pm$0.04 & WiggleZ (2011) \cite{Blake} \\
0.57 & 0.43$\pm$0.03 & BOSS CMASS (2012) \cite{Reid12} \\
0.6 & 0.43$\pm$0.04 & WiggleZ (2011) \cite{Blake} \\
0.78 & 0.38$\pm$0.04 & WiggleZ (2011) \cite{Blake} \\
\hline \hline
\end{tabular}
\caption{Data of $f\sigma_8$ measured from RSD
with the survey references. See also Table 1 of Ref. \cite{Samushia12}.}
\label{table:fs8data}
\end{table}
\end{center}

So, the main motivation of this paper is to investigate the effect of model parameter $c$ to $f\sigma_8(z)$ and to update our previous results by including the current observational data of RSD as well as SN Ia Union2.1, CMB and BAO on constraining the HDE model parameter space. 

This paper is structured as follows. In section \ref{sec:bgpe}, we give a very brief review of the HDE model where the radiation is included and the future event horizon is adopted as an IR cut-off. The scalar perturbation evolution equations for a spatially flat FRW Universe will also be presented. In section \ref{sec:method}, the constraint methodology and results will be presented. We give a summary in section \ref{sec:summary}.
 
\section{Background and Perturbation Evolution Equations} \label{sec:bgpe}

The energy density of the HDE is written as \cite{ref:holo2}
\begin{equation}
\rho_{h}=\frac{3c^2 M^2_{pl}}{R^2_{eh}}.\label{eq:EHHDE}
\end{equation}
The Friedmann equation for a spatially flat FRW universe reads
\begin{equation}
H^2=H^2_0\left(\Omega_{r0}a^{-4}+\Omega_{b0}a^{-3}+\Omega_{c0}a^{-3}\right)+\Omega_{h}H^2,\label{eq:EHFE}
\end{equation}
where $\Omega_{i}=\rho_i/3 M^2_{pl}H^2$ are dimensionless energy densities for radiation, baryon, cold dark matter and HDE respectively. Here the scale factor $a$ has been normalized to $a_0=1$ at present. Combining Eq. (\ref{eq:EH}) and Eq. (\ref{eq:EHHDE}), one obtains the differential equation for $\Omega_{h}$ \cite{ref:holoXu} 
\begin{equation}
\Omega_{h}'=-2\Omega_{h}\left(1-\Omega_{h}\right)\left(\frac{E'(x)}{E(x)}-\frac{\sqrt{\Omega_{h}}}{c}\right),\label{eq:diffeq}
\end{equation}
where $'$ denotes the derivative with respect to $x =\ln a$ and $E(x)=\sqrt{\Omega_{r0}e^{-2x}+\Omega_{b0}e^{-x}+\Omega_{c0}e^{-x}}$. This equation describes the evolution of dimensionless energy density of HDE with the initial condition $\Omega_{h0}=1-\Omega_{r0}-\Omega_{b0}-\Omega_{c0}$. Via the conservation equation of the HDE
$\dot{\rho_{h}}+3H(\rho_{h}+p_{h})=0$, one has the equation of state (EoS) of the HDE
\begin{equation}
w_{h}=-1-\frac{1}{3}\frac{d \ln \rho_{h}}{d \ln a}
=-\frac{1}{3}-\frac{2\sqrt{\Omega_{h}}}{3c},\label{eq:DEEOS}
\end{equation}
where the definition $w_{h}=p_{h}/\rho_{h}$ is used.

In this paper, the HDE is taken as a perfect fluid with the EoS (\ref{eq:DEEOS}), then in the synchronous gauge the perturbation equations of density contrast and velocity divergence for the HDE are written as
\begin{eqnarray}
\dot{\delta}_h&=&-(1+w_h)(\theta_h+\frac{\dot{h}}{2})-3\mathcal{H}(\frac{\delta p_{h}}{\delta \rho_{h}}-w_h)\delta_h,\label{eq:continue}\\
\dot{\theta}_h&=&-\mathcal{H}(1-3c^2_{s,ad})+\frac{\delta p_{h}/\delta \rho_{h}}{1+w_h}k^{2}\delta_h-k^{2}\sigma_h\label{eq:euler}
\end{eqnarray}
following the notations of Ma and Bertschinger \cite{ref:MB}, where the definition of the adiabatic sound speed
\begin{equation}
c^2_{s,ad}=\frac{\dot{p}_h}{\dot{\rho}_h}=w_h-\frac{\dot{w}_h}{3\mathcal{H}(1+w_h)}
\end{equation}
 is used. When the EoS of a pure barotropic fluid is negative, the imaginary adiabatic sound speed can cause instability of the perturbations. To overcome this problem, one can introduce an entropy perturbation and assume a positive or null effective speed of sound. Following the work of \cite{ref:Hu98}, the non adiabatic stress or entropy perturbation can be separated out
 \begin{equation}
 p_h\Gamma_h=\delta p_h-c^2_{s,ad}\delta \rho_h, \label{eq:entropyper}
\end{equation} 
which is gauge independent. In the rest frame of HDE, the entropy perturbation is specified as
 \begin{equation}
 w_h\Gamma_h=(c^2_{s,eff}-c^2_{s,ad})\delta^{rest}_{h},\label{eq:restframe}
 \end{equation}
where $c^2_{s,eff}$ is the effective speed of sound. Transforming into an arbitrary gauge 
 \begin{equation}
 \delta^{rest}_{h}=\delta_h+3\mathcal{H}(1+w_h)\frac{\theta_h}{k^2}\label{eq:gaugetrans}
 \end{equation}  
 gives a gauge-invariant form for the entropy perturbations. By using the Eqs (\ref{eq:entropyper},) (\ref{eq:restframe}) and (\ref{eq:gaugetrans}), one can recast Eqs. (\ref{eq:continue}), and (\ref{eq:euler}) into 
 \begin{widetext}
 \begin{eqnarray}
 \dot{\delta}_h&=&-(1+w_h)(\theta_h+\frac{\dot{h}}{2})+\frac{\dot{w}_h}{1+w_h}\delta_h-3\mathcal{H}(c^2_{s,eff}-c^2_{s,ad})\left[\delta_h+3\mathcal{H}(1+w_h)\frac{\theta_h}{k^2}\right]\\
\dot{\theta}_h
&=&-\mathcal{H}(1-3c^2_{s,eff})\theta_h+\frac{c^2_{s,eff}}{1+w_h}k^2\delta_h-k^2\sigma_h
 \end{eqnarray}
\end{widetext}
For the HDE, we assume the shear perturbation $\sigma_h=0$ and the adiabatic initial conditions. Actually, the effective speed of sound $c^2_{s,eff}$ is another freedom to describe the micro scale property of HDE in addition to the EoS \cite{Putter2010}. And, we should take it as another free model parameter. The sound speed determines the sound horizon of the fluid via the equation $l_s=c_{s,eff}/H$. The fluid can be smooth or cluster below or above the sound horizon $l_s$ respectively. If the sound speed is smaller, the perturbation of the fluid can be detectable on large scale. And in turn the clustering fluid can influence the growth of density perturbations of matter, large scale structure and evolving gravitational potential which generates the integrated Sachs-Wolfe (ISW) effects. However, the authors of \cite{Putter2010} have shown that current data can put no significant constraints on the value of the sound speed when dark energy is purely a recent phenomenon. For the HDE considered in this paper, it is related to the future event horizon and would not cluster. So we assume the effective speed of sound $c^2_{s,eff}=1$ in this work. 

\section{Methodology and Constraint Results} \label{sec:method}

In our previous work \cite{ref:holoXu}, we have used the SN Ia Union2, BAO and full information of CMB from WMAP-7yr to constrain the model parameter space, where the effects of model parameter $c$ to the CMB power spectrum were also discussed. In Refs. \cite{ref:holoXu}, we showed that large values of $c$ increase the tails of CMB power spectrum at large scale, i.e. $l<10$, through the integrated Sachs-Wolfe (ISW) effect. Here we will focus on its effects to the $f\sigma_8(z)$ caused by the different values of $c$. At first, we modify the {\bf CAMB} package which is the publicly available code\footnote{http://camb.info/.} for calculating the CMB power spectrum to include the HDE. We calculate the values of $\sigma_8$ at different redshift for the HDE model. We also write a subroutine to calculate the growth rate $f(z)$ for the HDE model. The growth rate can be obtained by solving the following differential equation \cite{ref:growthrate} 
\begin{eqnarray}
\frac{d^2g}{d\ln a^2}&+&\left[\frac{5}{2}+\frac{1}{2}\left(\Omega_k(a)-3w_{eff}(a)\Omega_{de}(a)\right)\right]\frac{dg}{d\ln a}\nonumber\\
&+&\left[2\Omega_{k}(a)+\frac{3}{2}\left(1-w_{eff}(a)\right)\Omega_{de}(a)\right]g=0,
\end{eqnarray}
where 
\begin{eqnarray}
g(a)&\equiv& \frac{D(a)}{a}=(1+z)D(z),\\
\Omega_{k}(a)&\equiv& \frac{\Omega_{k}H^2_0}{a^2H^2(a)},\\
\Omega_{de}(a)&\equiv&\frac{\Omega_{de}H^2_0}{a^{3[1+w_{eff}(a)]}H^2(a)},\\
w_{eff}(a)&\equiv& \frac{1}{\ln a}\int^{\ln a}_0 d\ln a' w(a').
\end{eqnarray}
Here $D(a)$ is the amplitude of the growing mode which connects to $f(a)$ via the relation $f\equiv d\ln D/d\ln a$. Finally, we can obtain the values of $f\sigma_8(z)$ at different redshift $z$. To investigate the effects of $c$ to $f\sigma_8(z)$, we borrow and fix the relevant cosmological values from our previous results obtained in \cite{ref:holoXu} but take the model parameter $c$ varying in a range. The evolution of $f\sigma_8(z)$ with respect to the redshift $z$ for different values of $c$ is shown in Figure \ref{fig:fsigms8c}. One can read off that the large values of $c$ decrease and increase the values of $f\sigma_8(z)$ at higher and lower redshifts respectively from the Figure \ref{fig:fsigms8c}. It clues that the $f\sigma_8$ data points favor the values of model parameter $c$ in a range of $[0.69,0.9]$. However, due to the sparseness and relative large error bars of the RSD data points, the current data sets of $f\sigma_8(z)$ may not give a much tight constraint to the model parameter space.  
   
\begin{widetext}
\begin{center}
\begin{figure}[htb]
\includegraphics[width=14cm]{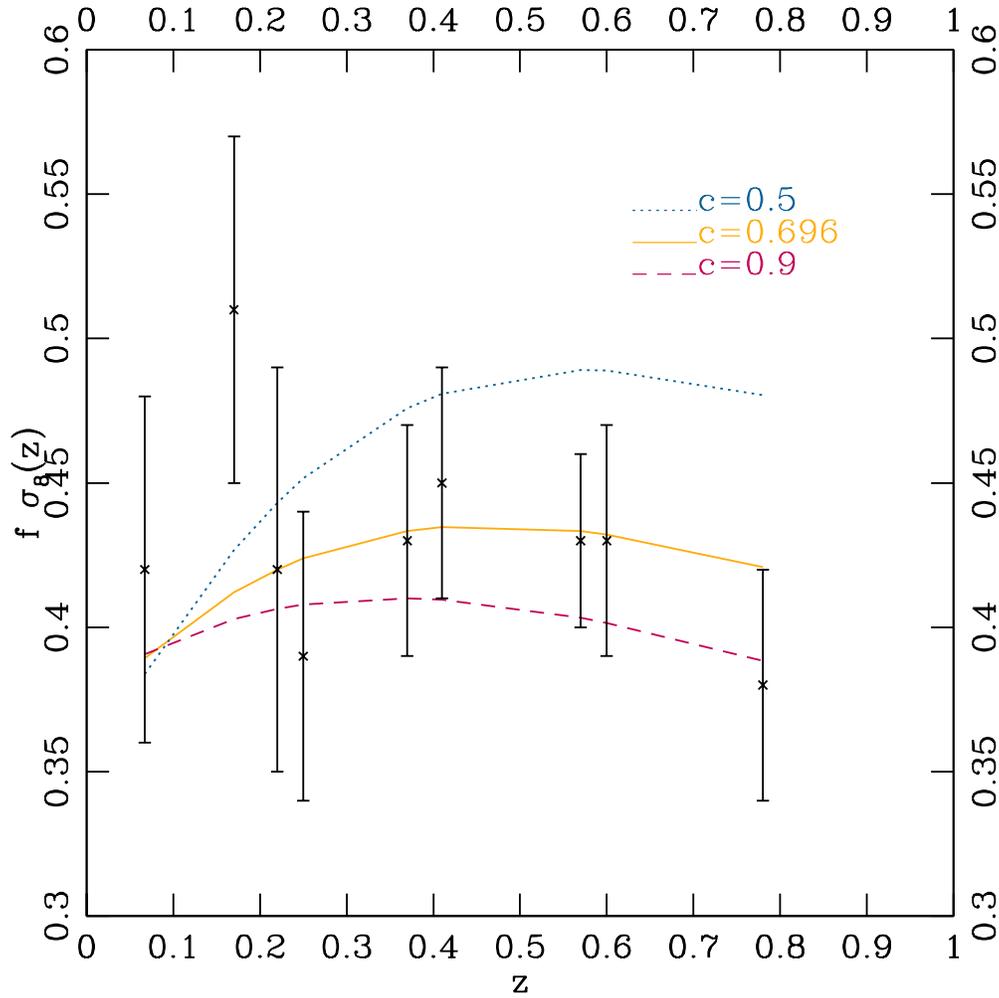}
\caption{The $f\sigma_8(z)$ v.s. the redshift $z$ for different values of model parameter $c$ (the red dashed line is for $c=0.9$, the orange thick line is for $c=0.696$ and the blue dotted line is for $c=0.5$ ), where the other relevant cosmological parameters are fixed to their mean values obtained in Ref. \cite{ref:holoXu}. Large values of $c$ decrease and increase the values of $f\sigma_8(z)$ at higher and lower redshifts respectively. The black lines with error bars denote the observed data points as listed in Table \ref{table:fs8data}.}\label{fig:fsigms8c}
\end{figure}
\end{center}
\end{widetext}

To obtain the model parameter space from currently available cosmic observations, we use the Markov Chain Monte Carlo (MCMC) method which is efficient in the case of more parameters case. We modified the publicly available {\bf cosmoMC} package\footnote{http://cosmologist.info/cosmomc/.} \cite{ref:MCMC} to include the likelihood coming from the $f\sigma_8(z)$. We adopted the $7$-dimensional parameter space
\begin{equation}
P\equiv\{\omega_{b},\omega_c, \Theta_{S},\tau, c,n_{s},\log[10^{10}A_{s}]\}
\end{equation}
the priors for the model parameters are summarized in Table \ref{tab:results}. Furthermore, the hard-coded prior on the comic age $10\text{Gyr}<t_{0}<\text{20Gyr}$ is also imposed. Also, the physical baryon density $\omega_{b}=\Omega_b h^2=0.022\pm0.002$ \cite{ref:bbn} from big bang nucleosynthesis and new Hubble constant $H_{0}=74.2\pm3.6\text{kms}^{-1}\text{Mpc}^{-1}$ \cite{ref:hubble} are adopted. The pivot scale of the initial scalar power spectrum $k_{s0}=0.05\text{Mpc}^{-1}$ is used in this paper.

The luminosity distance $d_L$ from SN Ia Uinon2.1 \cite{ref:Union21}, the angular diameter distance $D_A$ and CMB power spectra from WMAP-7yr are used to fix the background evolutions. For the details, please see Appendix \ref{sec:app}.

We ran eight chains on the {\it Computational Cluster for Cosmos} and stopped sampling when the worst e-values [the variance(mean)/mean(variance) of 1/2 chains] $R-1$ was of the order $0.01$. The global fitting results are summarized in the Table \ref{tab:results} and the Figure \ref{fig:contour}. Comparing to our previous result $c= 0.696_{-  0.0737-   0.132-  0.190}^{+   0.0736+   0.159+  0.264}$ \cite{ref:holoXu}, we find that SN Union2.1 favors large values of model parameter $c=0.737_{-    0.0826-    0.148-    0.202}^{+    0.0830+    0.196+    0.320}$. When the RSD $f\sigma_8(z)$ is included, the values of model parameter $c$ are increased to $c=0.750_{-    0.0999-    0.173-    0.226}^{+    0.0976+    0.215+    0.319}$ which confirms the analysis as shown in Figure \ref{fig:fsigms8c}.

\begin{widetext}
\begin{center}
\begingroup
\squeezetable
\begin{table}[tbh]
\begin{tabular}{cllclc}
\hline\hline Prameters & Priors & Mean with errors without $f\sigma_8$ & Best fit without $f\sigma_8$ & Mean with errors with $f\sigma_8$  & Best fit with $f\sigma_8$\\ \hline
$\Omega_b h^2$ & $[0.005,0.1]$ & $    0.0227_{-    0.000524-    0.00100-    0.00151}^{+    0.000517+    0.00104+    0.00165}$ & $0.226$ & $    0.0226_{-    0.000549-    0.00110-    0.00153}^{+    0.000542+    0.00117+    0.00164}$ & $0.0226$\\
$\Omega_{DM} h^2$ & $[0.01,0.99]$ & $    0.110_{-    0.00440-    0.00863-    0.0122}^{+    0.00446+    0.00888+    0.0135}$ & $0.111$ & $    0.110_{-    0.00466-    0.00992-    0.0127}^{+    0.00478+    0.00883+    0.0145}$ & $0.110$\\
$\theta$ & $[0.5,10]$ & $    1.0395_{-    0.00261-    0.00505-    0.00733}^{+    0.00264+    0.00512+    0.00781}$ & $1.0401$ & $    1.0394_{-    0.00271-    0.00530-    0.00791}^{+    0.00260+    0.00532+    0.00758}$ & $1.0392$\\
$\tau$ & $[0.01,0.8]$ & $    0.0896_{-    0.00759-    0.0233-    0.0368}^{+    0.00674+    0.0255+    0.0447}$ & $0.0846$ & $    0.0888_{-    0.00724-    0.0236-    0.0388}^{+    0.00628+    0.0250+    0.0466}$ & $0.0903$ \\
$c$ & $[0.1,1.5]$ & $    0.737_{-    0.0826-    0.148-    0.202}^{+    0.0830+    0.196+    0.320}$ & $0.713$ & $    0.750_{-    0.0999-    0.173-    0.226}^{+    0.0976+    0.215+    0.319}$ & $0.733$\\
$n_s$ & $0.5,1.5$ & $    0.972_{-    0.0124-    0.0243-    0.0370}^{+    0.0126+    0.0267+    0.0407}$ & $0.970$ & $    0.972_{-    0.0131-    0.0259-    0.0403}^{+    0.0132+    0.0275+    0.0436}$ & $0.970$\\
$\log[10^{10} A_s]$ & $[2.4,4]$ & $    3.0795_{-    0.0341-    0.0669-    0.0940}^{+    0.0343+    0.0690+    0.108}$ & $3.0730$ & $    3.0766_{-    0.0366-    0.0690-    0.100}^{+    0.0357+    0.0762+    0.114}$ & $3.0795$\\
\hline
$\Omega_h$ & - & $    0.719_{-    0.0176-    0.0375-    0.0592}^{+    0.0183+    0.0346+    0.0510}$ & $0.719$ & $    0.717_{-    0.0174-    0.0369-    0.0587}^{+    0.0171+    0.0321+    0.0461}$ & $0.717$\\
$Age/Gyr$ & - & $   13.901_{-    0.109-    0.216-    0.332}^{+    0.109+    0.220+    0.314}$ & $13.886$ & $   13.916_{-    0.114-    0.221-    0.366}^{+    0.113+    0.240+    0.354}$ & $13.923$\\
$\Omega_m$ & - & $    0.281_{-    0.0183-    0.0346-    0.0508}^{+    0.0176+    0.0375+    0.0595}$ & $0.281$ & $    0.283_{-    0.0171-    0.0320-    0.0460}^{+    0.0174+    0.0369+    0.0587}$ & $0.283$\\
$\sigma_8$ & - & - & - & $    0.763_{-    0.0465-    0.0826-    0.108}^{+    0.0477+    0.0910+    0.120}$ & $0.766$\\
$z_{re}$ & - & $   10.647_{-    1.219-    2.273-    3.401}^{+    1.186+    2.409+    3.730}$ & $10.302$ & $   10.578_{-    1.189-    2.442-    3.629}^{+    1.189+    2.335+    3.794}$& $10.770$\\
$H_0$ & - & $   68.787_{-    1.839-    3.680-    5.757}^{+    1.836+    3.847+    5.602}$ & $68.927$ & $   68.414_{-    1.904-    3.795-    5.216}^{+    1.885+    3.780+    5.706}$ & $68.451$\\
\hline\hline
\end{tabular}
\caption{The mean values with $1,2,3\sigma$ errors and the best fit values of the model parameters and derived cosmological parameters, where the WMAP $7$-year, SN Union2.1, BAO and RSD $f\sigma_8$ data sets are used.}\label{tab:results}
\end{table}
\endgroup
\end{center}
\end{widetext}

\begin{widetext}
\begin{center}
\begin{figure}[tbh]
\includegraphics[width=18cm]{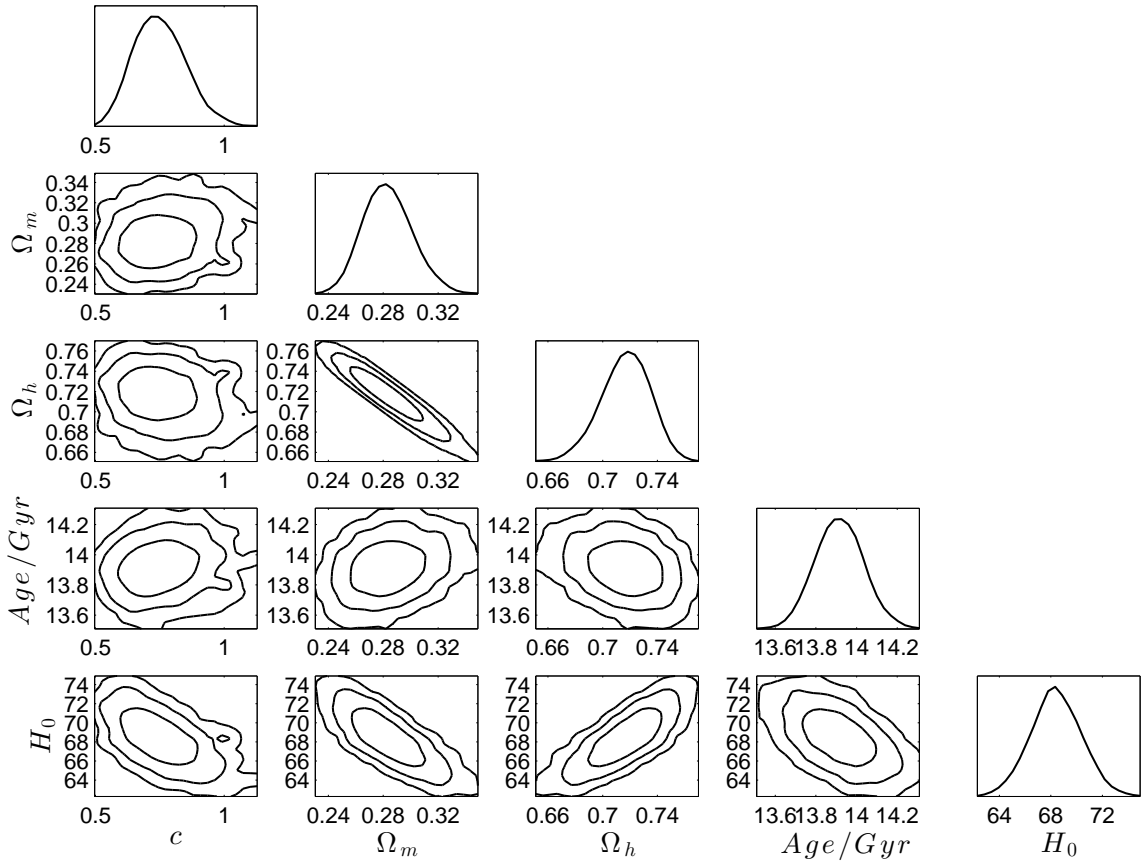}
\caption{The 1D marginalized distribution on individual parameters and 2D contours  with $68\%$ C.L., $95\%$ C.L. and  $99.7\%$ C.L. by using CMB+BAO+SN+RSD data points.}\label{fig:contour}
\end{figure}
\end{center}
\end{widetext}

To show the effects of RSD data points $f\sigma_8(z)$ to constrain the model parameters space, the 2D contour for model parameter $\Omega_m-c$ is also plotted in Figure \ref{fig:compcont}. From this figure, one can read that the region of $\Omega_m$ is shrunk when $f\sigma_8(z)$ data points are employed. But in this case, the $1,2,3\sigma$ regions of $c$ are enlarged. And the 2D contour diagram moves little to the top right corner direction when $f\sigma_8(z)$ data points are included.   
\begin{widetext}
\begin{center}
\begin{figure}[htb]
\includegraphics[width=14cm]{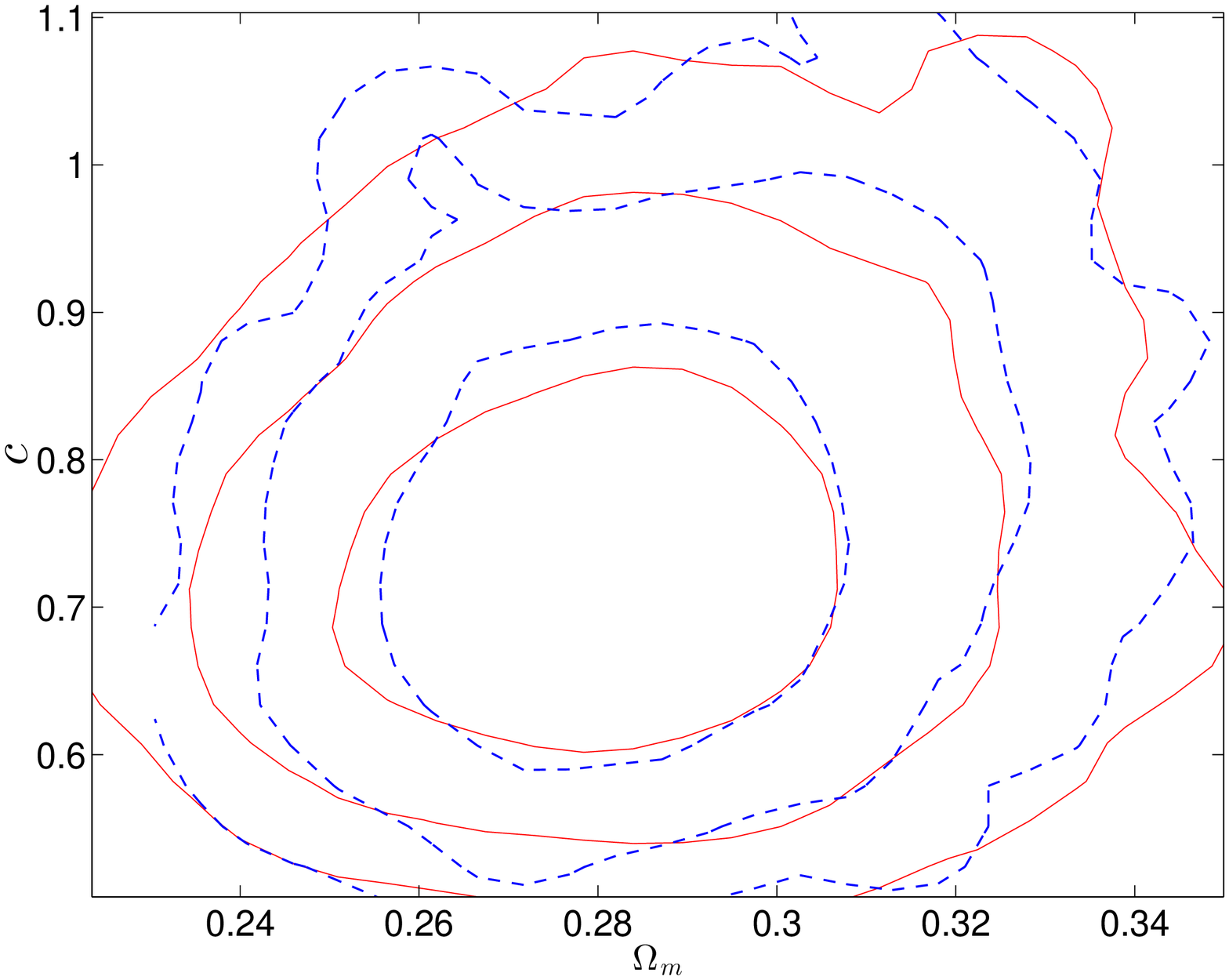}
\caption{The 2D contours with $68\%$ C.L. , $95\%$ C.L. and  $99.7\%$ C.L. for model parameter $\Omega_m-c$, where the red solid line is for CMB+BAO+SN, and the blue dashed line is for CMB+BAO+SN+RSD.}\label{fig:compcont}
\end{figure}
\end{center}
\end{widetext}

With the mean values listed in the Table \ref{tab:results} for the case of SN+BAO+CMB+RSD, we plotted the evolutions of the EoS of HDE with respect to the redshift $z$ in Figure \ref{fig:wz}, where the shadows denote the $1,2,3\sigma$ regions from the dark to the light respectively. For calculating the $1\sigma$ region, we consider the propagation of the errors for $w(z)$ and marginalize the other irrelevant model parameters by the Fisher matrix analysis \cite{ref:NRP,ref:Alam}. If the other 
irrelevant model parameters are not marginalized, the error bars will be underestimated. The errors are calculated by using the covariance matrix $C_{ij}$ of the fitting model parameters which is an output of {\bf cosmoMC}. The errors for a function $f=f(\theta)$ in terms of the variables $\theta$ are given via the formula \cite{ref:Alam,ref:Nesseris2005,ref:Wang2010}
\begin{equation}
\sigma^2_f=\sum^n_i\left(\frac{\partial f}{\partial \theta_i}\right)^2C_{ii}+2\sum^n_i\sum^n_{j=i+1}\left(\frac{\partial f}{\partial \theta_i}\right)\left(\frac{\partial f}{\partial \theta_j}\right)C_{ij}
\end{equation}
 where $n$ is the number of the variables. In our case, $f$ would be the EoS $w(z;\theta_i)$ for HDE. And the variables $\theta_i$ are $(\Omega_b h^2,\Omega_c h^2,c)$ for the HDE model. The corresponding $1\sigma$ errors for $w(z)$ are given by
 \begin{equation}
 w_{1\sigma}(z)=w(z)|_{\theta=\bar{\theta}}\pm \sigma_w,
 \end{equation}
 where $\bar{\theta}$ are the mean values of the constrained model parameters. For a relative large values of $c$, the HDE behaves like quintessence at present ($w_h|_{z=0}=-0.971\pm0.0777$ with $1\sigma$ error). In $2\sigma$ regions, it still has broad space to behave like phantom even in the future. But in $3\sigma$ region, it has the possiblity to behave like quintessence. So based on this result, we still do not know our Universe will be terminated by a cosmic doomsday or not in $3\sigma$ region.  
 
\begin{widetext}
\begin{center}
\begin{figure}[htb]
\includegraphics[width=14cm]{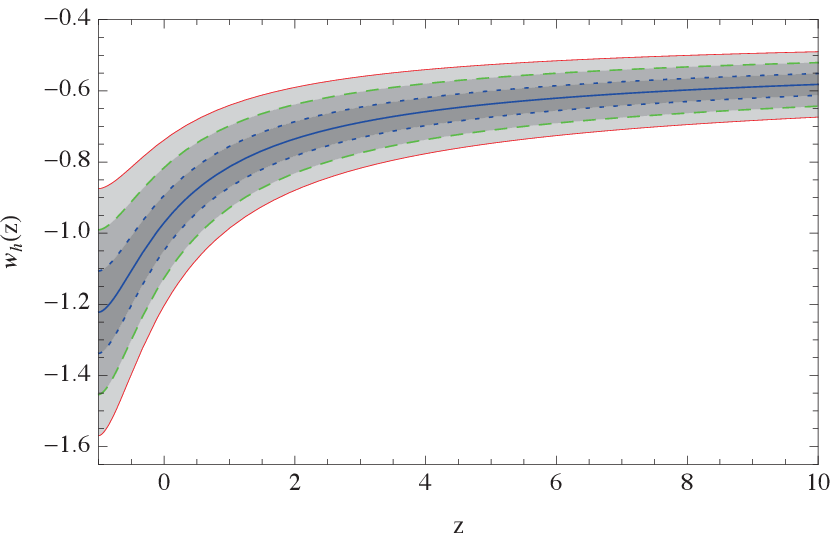}
\caption{The evolution of EoS for HDE with $1,2,3\sigma$ shadow regions, where the mean values of the relevant model parameter are adopted as listed in the Table \ref{tab:results} for the case CMB+BAO+SN+RSD.}\label{fig:wz}
\end{figure}
\end{center}
\end{widetext}

\section{Summary} \label{sec:summary}

In this paper, we updated our previous results obtained in Ref. \cite{ref:holoXu} with the replacement of SN Union2 by SN Union2.1 and with the addition of RSD data points of $f\sigma_8(z)$. We showed the effects of model parameter $c$ to $f\sigma_8(z)$ by fixing the other relevant model parameters and found out that RSD $f\sigma_8(z)$ data points favor larger values of $c$. But due to the sparseness and relative large error bars of the RSD data points, the current data sets of $f\sigma_8(z)$ cannot give a much tight constraint to the model parameter $c$. A global fitting to the HDE model was performed by combining the full information of CMB from WMAP-7yr, BAO, SN Union2.1, with and without RSD $f\sigma_8(z)$ data sets via the MCMC method. The results show that RSD data points $f\sigma_8(z)$ can shrink the model parameter space $\Omega_m$ efficiently as shown in Figure \ref{fig:compcont} but cannot constrain the model parameter $c$ very well.  When the RSD $f\sigma_8(z)$ data points are added, the 2D contour diagram moves little to the top right corner direction on the 2D $\Omega_m-c$ plane as shown in Figure \ref{fig:compcont}. It means that the RSD $f\sigma_8(z)$ data points favor larger values of $c$ and $\Omega_m$. It confirms our previous analysis as shown in Figure \ref{fig:fsigms8c}.  

To show the evolution of the EoS with errors for HDE with respect to the redshift $z$, we should margnialize the other irrelevant model parameters. If not the error bars will be under estimated. We marginalized the other irrelevant model parameters by the Fisher matrix analysis. And the evolution of the EoS for HDE in $3\sigma$ region was plotted in Figure \ref{fig:wz} by adopting the mean values as shown in Table \ref{tab:results}. In this figure one can see that HDE behaves like quintessence at present ($w_h|_{z=0}=-0.971\pm0.0777$ with $1\sigma$ error). In $2\sigma$ region, it has a wide region to behave like phantom. But in $3\sigma$ region, it has possiblities to behave like quintessence. Then one still cannot conclude whether the future Universe will terminated by a cosmic doomsday or not in $3\sigma$ region.

\acknowledgements{The author thanks an anonymous referee for helpful improvement of this paper. L. Xu's work is supported in part by NSFC under the Grants No. 11275035 and "the Fundamental Research Funds for the Central Universities" under the Grants No. DUT13LK01.}

\appendix

\section{SN Ia Union2.1, BAO and CMB}\label{sec:app}

For the SN Ia, the Uinon2.1 \cite{ref:Union21} data sets will be used in this paper. The distance modulus $\mu(z)$ is defined as
\begin{equation}
\mu_{th}(z)=5\log_{10}[\bar{d}_{L}(z)]+\mu_{0},
\end{equation}
where $\bar{d}_L(z)$ is the Hubble-free luminosity distance $H_0
d_L(z)/c=H_0 d_A(z)(1+z)^2/c$, with $H_0$ the Hubble constant, and
$\mu_0\equiv42.38-5\log_{10}h$ through the re-normalized quantity
$h$ as $H_0=100 h~{\rm km ~s}^{-1} {\rm Mpc}^{-1}$. Where $d_L(z)$
is defined as
\begin{eqnarray}
d_L(z)&=&(1+z)r(z)\\
r(z)&=&\frac{c}{H_0\sqrt{|\Omega_{k}|}}{\rm
sinn}\left[\sqrt{|\Omega_{k}|}\int^z_0\frac{dz'}{E(z')}\right]
\end{eqnarray}
where $E^2(z)=H^2(z)/H^2_0$. Additionally, the observed distance
moduli $\mu_{obs}(z_i)$ of SN Ia at $z_i$ are
\begin{equation}
\mu_{obs}(z_i) = m_{obs}(z_i)-M,
\end{equation}
where $M$ is their absolute magnitudes.

For the SN Ia dataset, the best fit values of the parameters $p_s$
can be determined by a likelihood analysis, based on the calculation
of
\begin{eqnarray}
\chi^2(P,M^{\prime})&\equiv& \sum_{SN}\frac{\left\{
\mu_{obs}(z_i)-\mu_{th}(P,z_i)\right\}^2} {\sigma_i^2}  \nonumber\\
&=&\sum_{SN}\frac{\left\{ 5 \log_{10}[\bar{d}_L(P,z_i)] -
m_{obs}(z_i) + M^{\prime} \right\}^2} {\sigma_i^2}, \label{eq:chi2}
\end{eqnarray}
where $M^{\prime}\equiv\mu_0+M$ is a nuisance parameter which
includes the absolute magnitude and the parameter $h$. The nuisance parameter $M^{\prime}$ can be marginalized over
analytically \cite{ref:SNchi2} as
\begin{equation}
\bar{\chi}^2(P) = -2 \ln \int_{-\infty}^{+\infty}\exp \left[
-\frac{1}{2} \chi^2(P,M^{\prime}) \right] dM^{\prime},\nonumber
\label{eq:chi2marg}
\end{equation}
resulting to
\begin{equation}
\bar{\chi}^2 =  A - \frac{B^2}{C} + \ln \left( \frac{C}{2\pi}\right), \label{eq:chi2mar}
\end{equation}
with
\begin{widetext}
\begin{eqnarray}
A&=&\sum_{i,j}^{SN}\left\{5\log_{10}
[\bar{d}_L(P,z_i)]-m_{obs}(z_i)\right\}\cdot {\rm
Cov}^{-1}_{ij}\cdot \left\{5\log_{10}
[\bar{d}_L(P,z_j)]-m_{obs}(z_j)\right\},\nonumber\\
B&=&\sum_i^{SN} {\rm Cov}^{-1}_{ij}\cdot \left\{5\log_{10}
[\bar{d}_L(P,z_j)]-m_{obs}(z_j)\right\},\nonumber \\
C&=&\sum_i^{SN} {\rm Cov}^{-1}_{ii},\label{eq:SNsyserror}
\end{eqnarray}
\end{widetext}
where ${\rm Cov}^{-1}_{ij}$ is the inverse of covariance matrix with
or without systematic errors. One can find the details in Ref.
\cite{ref:Union21} and the web site
\footnote{http://supernova.lbl.gov/Union/.} where the covariance
matrix with or without systematic errors are included. Relation
(\ref{eq:chi2}) has a minimum at the nuisance parameter value
$M^{\prime}=B/C$, which contains information of the values of $h$
and $M$. Therefore, one can extract the values of $h$ and $M$
provided the knowledge of one of them. Finally, the expression
\begin{equation}
\chi^2_{SN}(P,B/C)=A-(B^2/C),\label{eq:chi2SN}
\end{equation}
which coincides to Eq. (\ref{eq:chi2mar}) up to a constant, is often
used in the likelihood analysis \cite{ref:smallomega,ref:SNchi2}.
Thus in this case the results will not be affected by a flat
$M^{\prime}$ distribution. It worths noting that the results will be
different with or without the systematic errors. In this work, all
results are obtained with systematic errors.

For BAO data sets, we used the observational results $d^{obs}_z$ from SDSS DR7 \cite{SDSSDR7} and $A(z)$ from WiggleZ \cite{Blake}. The observed values of $d^{obs}_z$ are gathered in Table \ref{tab:SDSSDR7}.
\begin{center}
\begin{table}[tbh]
    \begin{tabular}{ccc}
    \hline\hline
      $z$ & $d^{obs}_z$ & survey and reference\\
        \hline
	0.20 & $0.1905 \pm 0.0061$ & SDSS \cite{SDSSDR7} \\
	0.35 & $0.1097 \pm 0.0036$ & SDSS \cite{SDSSDR7} \\
	\hline\hline
      \end{tabular}
  \caption{The $d^{obs}_z$ from SDSS DR7 \cite{SDSSDR7} .}
\end{table}\label{tab:SDSSDR7}
 \end{center}
  where $d_z\equiv r_s(z_d)/D_V(z)$, $r_s(z_d)$ is the comoving sound horizon at the baryon drag epoch, $D_V(z)\equiv [(1+z)^2D^2_Acz/H(z)]^{1/3}$ \cite{ref:DV1,ref:DV2}. Here $D_A(z)$ the angular diameter distance which is defined as
\begin{equation}
D_A(z)=\frac{r(z)}{1+z}.
\end{equation}
 For the SDSS DR7 data points, the $\chi^2_{SDSS}(P)$ is given as
 \begin{equation}
\chi^2_{SDSS}(P)=\sum^{SDSS}_{i,j}(d^{th}_i(P)-d^{obs}_i)\cdot C^{-1}_{ij}\cdot (d^{th}_j(P)-d^{obs}_j)
\end{equation}
where $C^{-1}$ is the inverse covariance matrix 
\begin{equation}
C^{-1} =
\begin{pmatrix}
30124 & -17227 \\
-17227 & 86977
\end{pmatrix}
\end{equation}

 To calculate $r_{s}(z_{d})$, one needs to know the redshift $z_d$ at decoupling epoch and its corresponding sound horizon. We obtain the baryon drag epoch redshift $z_d$ numerically from the following integration \cite{ref:Hamann}
\begin{eqnarray}
\tau(\eta_d)&\equiv& \int_{\eta}^{\eta_0}d\eta'\dot{\tau}_d\nonumber\\
&=&\int_0^{z_d}dz\frac{d\eta}{da}\frac{x_e(z)\sigma_T}{R}=1
\end{eqnarray}   
where $R=3\rho_{b}/4\rho_{\gamma}$, $\sigma_T$ is the Thomson cross-section and $x_e(z)$ is the fraction of free electrons. Then the sound horizon is
\begin{equation}
r_{s}(z_{d})=\int_{0}^{\eta(z_{d})}d\eta c_{s}(1+z).
\end{equation}   
where $c_s=1/\sqrt{3(1+R)}$ is the sound speed. Also, to obtain unbiased parameter and error estimates, we use the substitution \cite{ref:Hamann}
\begin{equation}
d_z\rightarrow d_z\frac{\hat{r}_s(\tilde{z}_d)}{\hat{r}_s(z_d)}r_s(z_d),
\end{equation}
where $d_z=r_s(\tilde{z}_d)/D_V(z)$, $\hat{r}_s$ is evaluated for the fiducial cosmology of Ref. \cite{SDSSDR7}, and $\tilde{z}_d$ is redshift of drag epoch obtained by using the fitting formula \cite{ref:EH} for the fiducial cosmology

For WiggleZ data points, one calculates acoustic parameter $A(z)$ introduced by Eisenstein et al. \cite{ref:DV1}
\begin{equation}
A(z)\equiv \frac{100 D_V(z)\sqrt{\Omega_m h^2}}{cz}.
\end{equation}
The observed values of $A(z)$ are gathered in Table \ref{tab:AZ}
\begin{center}
\begin{table}[tbh]
    \begin{tabular}{ccc}
    \hline\hline
      $z$ & $A^{obs}(z)$ & survey and reference\\
        \hline
	0.44 & $0.474 \pm 0.034$ & WiggleZ \cite{Blake} \\
	0.60 & $0.442 \pm 0.020$ & WiggleZ \cite{Blake} \\
	0.73 & $0.424 \pm 0.021$ & WiggleZ \cite{Blake} \\
	\hline\hline
      \end{tabular}
  \caption{The $A(z)$ from WiggleZ \cite{Blake} .}
\end{table}\label{tab:AZ}
 \end{center}
 The corresponding $\chi^2_{WiggleZ}$ is given as
  \begin{eqnarray}
\chi^2_{WiggleZ}(P)=\left.\sum^{WiggleZ}_{i,j}(A^{th}(P,z_i)-A^{obs}(z_i))\right.\nonumber\\
\left.\cdot C^{-1}_{ij}\cdot (A^{th}(P,z_j)-A^{obs}(z_j))\right.
\end{eqnarray}
where $C^{-1}$ is the inverse covariance matrix 
\begin{equation}
C^{-1} =
\begin{pmatrix}
1040.3 & -807.5 & 36.8 \\
-807.5 & 3720.3 & -1551.9\\
336.8 & -1551.9 & 2914.9
\end{pmatrix}
\end{equation}

 Then the total $\chi^2_{BAO}$ from BAO is written as
\begin{equation}
\chi^2_{BAO}(P)=\chi^2_{SDSS}(P)+\chi^2_{WiggleZ}(P).
\end{equation}

For the $f\sigma_8(z)$, the $\chi^2_{f\sigma_8}(P)$ is given 
\begin{equation}
\chi^2_{f\sigma_8}(P)=\sum_i^{f\sigma_8}\frac{(f\sigma^{th}_8(P,z_i)-f\sigma^{obs}_8(z_i))^2}{\sigma^2_{f\sigma_8i}}.
\end{equation}

For CMB data set, the temperature power spectrum from WMAP $7$-year data\footnote{http://lambda.gsfc.nasa.gov/product/map/current/likelihood\_get.cfm.} \cite{ref:wmap7} are employed.

Then one has the total likelihood $\mathcal{L} \propto e^{-\chi^{2}/2}$, where $\chi^{2}$ is given as
\begin{equation}
\chi^{2}(P)=\chi^{2}_{SN}(P)+\chi^{2}_{BAO}(P)+\chi^2_{f\sigma_8}(P)+\chi^{2}_{CMB}(P),
\end{equation}
which is used to get the distribution of the model parameter space.

\end{document}